\documentclass[letterpaper]{article} 
\usepackage{aaai23}  
\usepackage{times}  
\usepackage{helvet}  
\usepackage{courier}  
\usepackage[hyphens]{url}  
\usepackage{graphicx} 
\usepackage{natbib}  
\usepackage{caption} 
\usepackage{url}
\usepackage{graphicx}
\usepackage{amsmath}
\usepackage{booktabs}
\usepackage{bbm}
\usepackage{graphicx}
\usepackage{newfloat}
\usepackage{listings}
\usepackage{color}
\usepackage{amssymb}
\usepackage{multirow}
\usepackage{multicol}
\usepackage{lineno,hyperref}
\modulolinenumbers[5]

\usepackage{amsmath,amssymb,amsfonts}
\usepackage{algorithmic}
\usepackage{algorithm}
\usepackage{graphicx}
\usepackage{textcomp}
\usepackage{colortbl}
\usepackage{xcolor}
\usepackage{color}
\usepackage{listings}
\usepackage{xcolor}
\usepackage{newfloat}
\usepackage{listings}
\DeclareCaptionStyle{ruled}{labelfont=normalfont,labelsep=colon,strut=off} 
\floatstyle{ruled}
\newfloat{listing}{tb}{lst}{}
\floatname{listing}{Listing}
\usepackage{subfigure} 
\usepackage{graphicx}
\newtheorem{definition}{Definition}
\usepackage{amsmath}
\usepackage{cite}

\usepackage{bibentry}

\begin{document}

\title{Finding Similar Exercises in Retrieval Manner}

\author{Tongwen Huang, Xihua Li, Chao Yi, Xuemin Zhao, Yunbo Cao}
\affiliations{Tencent Inc. China \\
\{towanhuang,xihuali,chyi,xueminzhao,yunbocao\}@tencent.com
}

\maketitle

\begin{abstract}
 When students make a mistake in an exercise, they can consolidate it by ``similar exercises'' which have the same concepts, purposes and methods. Commonly, for a certain subject and study stage, the size of the exercise bank is in the range of millions to even tens of millions, how to find similar exercises for a given exercise becomes a crucial technical problem. Generally, we can assign a variety of explicit labels to the exercise, and then query through the labels, but the label annotation is time-consuming, laborious and costly, with limited precision and granularity, so it is not feasible. In practice, we define ``similar exercises'' as a retrieval process of finding a set of similar exercises based on recall, ranking and re-rank procedures, called the \textbf{FSE} problem (Finding similar exercises). In other papers, FSE focuses on ranking method, this paper will comprehensively introduce recall, ranking, re-rank, and define similar exercise more accurately. Furthermore, comprehensive representation of the semantic information of exercises was obtained through representation learning. In addition to the reasonable architecture, we also explore what kind of tasks are more conducive to the learning of exercise semantic information from pre-training and supervised learning. It is difficult to annotate similar exercises and the annotation consistency among experts is low. Therefore this paper also provides solutions to solve the problem of low-quality annotated data. Compared with other methods, this paper has obvious advantages in both architecture rationality and algorithm precision, which now serves the daily teaching of hundreds of schools.
\end{abstract}

\section{INTRODUCTION}
\label{introduction}

\begin{figure}[!hbt]
  \centering
\includegraphics[width=1.0\linewidth]{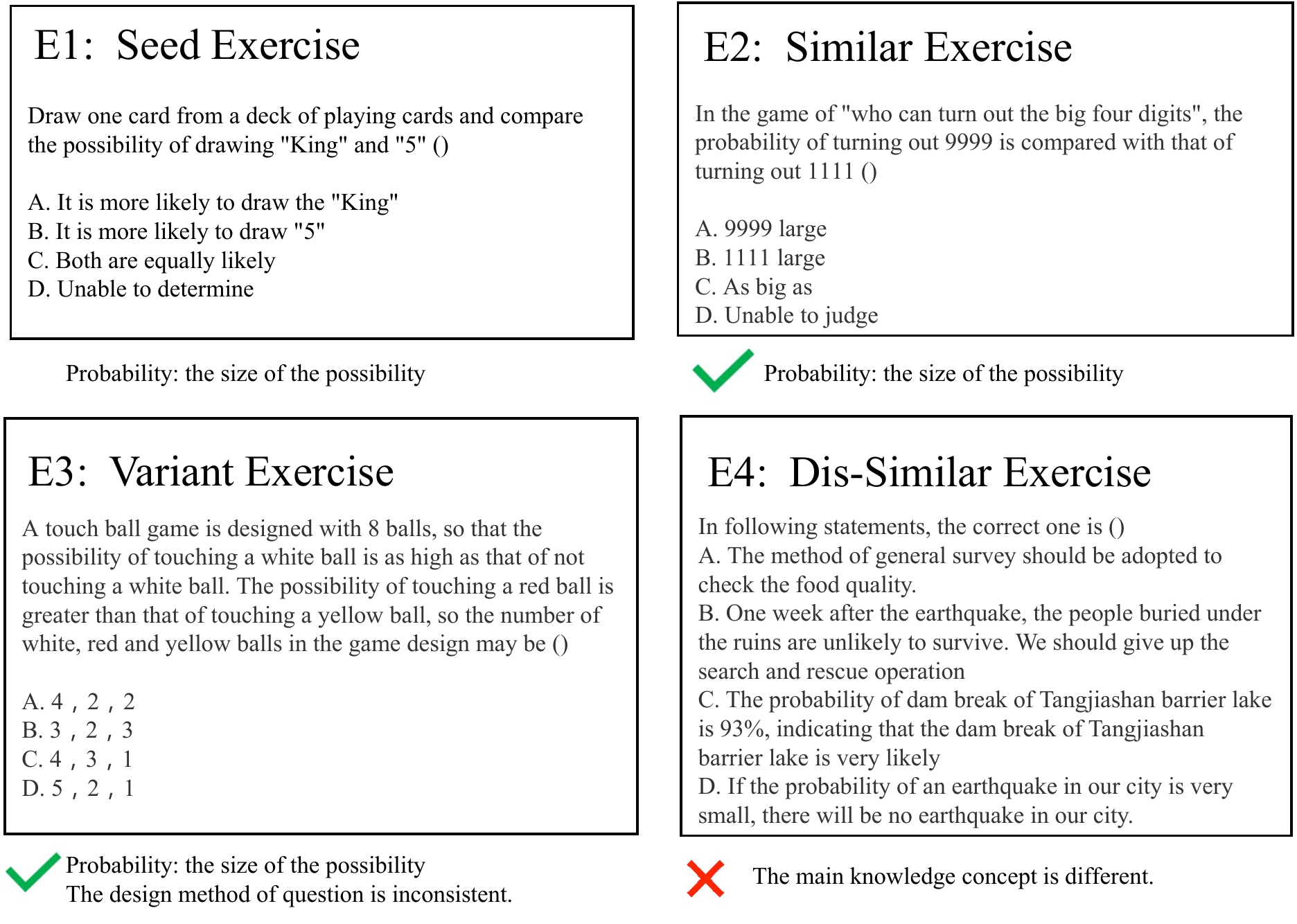}
  \caption{Examples of similar exercises. E1, E2, E3 and E4, the generalized similar exercise of E1 includes E2 and E3.}
  \label{fig:fmath}
  \hfill 
\end{figure}

With the rapid informatization of education, education artificial intelligence(EduAI) focuses on applying methods of artificial intelligence to benefit education tasks, which can help to improve the performance of the students and the teaching quality of the teachers. Intelligent test paper generation and consolidation exercises are important exercise-based applications. The teachers choose the similar exercises to replace their not satisfied ones when they generate the test paper and the students obtain some personalized consolidation exercises to improve their performances. The main technique behind is how to match the exercises, known as the finding similar exercises(FSE) problem. The generalized similar exercise includes \textit{similar exercise} and \textit{variant exercise}, which is concerned with finding an exercise having similar logic and relationships by understanding the semantics of exercise. As in Figure \ref{fig:fmath}, exercise E1 and its similar ones E2 and E3 share the same purpose of assessing the probability, while the dissimilar one E4 has different purpose. 

\begin{figure}[!hbt]
  \centering
\includegraphics[width=1.0\linewidth]{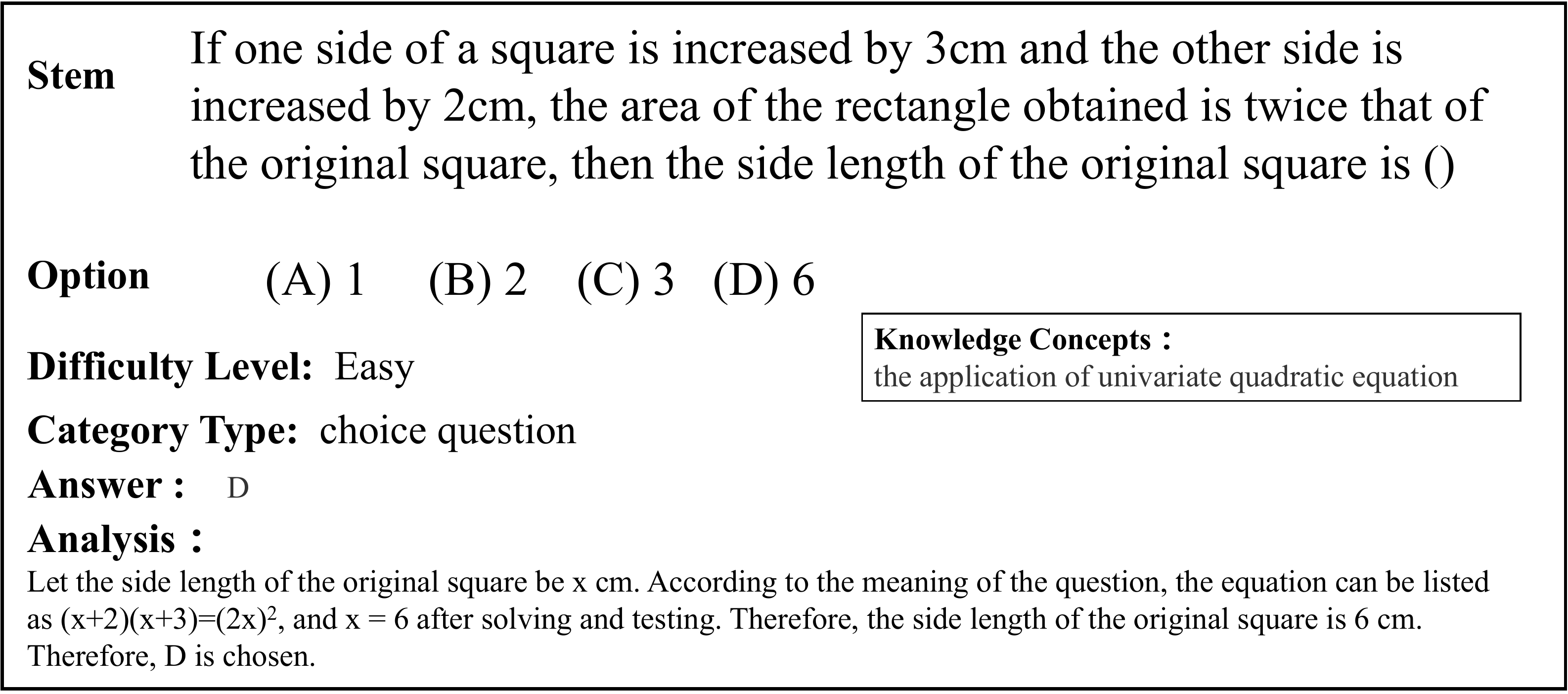}
  \caption{Contents of multi-modal exercise.}
  \label{fig:exe}
  \hfill 
\end{figure}

However, there are still many challenges to find similar exercises. First, how to present an exercise? Exercises are composed of multi-modals, including text, formulas, images (such as geometry ones), tables (such as statistics ones) and other information as shown in Figure \ref{fig:exe}. Second, how to obtain high-quality similar exercises? Is it that the more similar the text of the exercises, the more similar it is? Obviously not, the exercise with high text similarity may not be suitable, because the same exercise has no value to enhance students' learning effect. Third, owing to the difficulty and easy-confused of similar exercises, there is a lot of label noises, how to learn a model in noise dataset is a big challenge.Finally, previous works 
 \cite{liu2018finding,feng2019learning,yin2019quesnet} treat FSE as an exercise matching algorithm, we design an overall personalized exercises retrieval system instead. What is a reasonable architecture for finding similar exercises?  

In this paper, finding similar exercises (FSE) is modeled as semantic similarity retrieval, contributions listed:
\\\textbf{1.} The paper utilizes recall, ranking and re-rank framework to conduct FSE as conventional semantic retrieval. In the recall stage, possible candidates are obtained from the exercise bank (size of millions to tens of millions). In the ranking stage, similarity scores are obtained based on the semantic representation of exercises. In the re-rank stage, special strategies of education are utilized, such as learning stage filtering, personalization, etc.
\\\textbf{2.} According to the multi-modal characteristics of exercise, a multi-modal framework was proposed to better encode the structural information. In particular, multi-modal pre-training was utilized to improve the effect of representation learning, in order to overcoming the lack of labeled data.
\\\textbf{3.} For the ranking model, we found that the new tasks that were strongly related to the semantics of exercise (eg. exercise solving process) significantly improved the effect of FSE.
\\\textbf{4.} Because the consistency rate of annotated data is poor, confidence learning is introduced to improve the model.

\section{RELATED WORK}
\label{related_work}

\subsection{Exercise Representation Learning}
Inspired by the successful applications of representation learning in NLP and CV, various exercise representation learning methods have been proposed. Pardos \cite{pardos2017imputing} trained skip-grams model for representing exercises and their relationships to one another in Euclidian space with sequences of exercise IDs encountered by students. Liu \cite{liu2018finding} designed an Attention based Long Short-Term Memory (Attention-based LSTM) network to learn a unified semantic representation of each exercise by handling its heterogeneous materials in a multi-modal way. Feng \cite{feng2019learning} used SBERT to capture mathematical logic and relationships to learn appropriate exercise vector representation. QuesNet \cite{yin2019quesnet} is the most similar work with ours in the recall module. Both of us adopt the pretraining then fine-tuning approach. Differences exist in the design of the pretraining tasks and strategy and the scale of used pretraining data. Our pretraining strategy is more exercise representation specified and FSE task targeted.

\subsection{Multi-Task Learning}

Multi-task learning (MTL) is a field of machine learning where multiple tasks are learned in parallel while using a shared representation. Multi-task learning based deep learning has been applied to many real-world applications, such as natural language processing \cite{Liu2019multi,peng2016improving,wang2013joint}, computer vision \cite{lee2019multi}, recommendation system \cite{jacobs1991adaptive,ma2018modeling}. MT-DNN \cite{Liu2019multi} proposes a  BERT based multi-task hard sharing learning framework which combines different tasks such as single-sentence classification, pairwise text classification, text similarity, and relevance ranking. By sharing the representation of the bottom BERT and top layer has task-specific parameters,  MT-DNN obtains better performance improvement on GLUE's tasks. In addition, there are some studies about the information fusion of multi-task learning. MOE \cite{jacobs1991adaptive} and MMOE \cite{ma2018modeling} proposed to share some experts at the bottom layer and combined experts through a gating network. However, few previous works applied MTL in education field. MTL is particularly helpful in education field where labeled datasets are hard to collect owing to the strong expertise.

\subsection{Contrastive Learning}
The main idea of contrastive learning (CL) is to learn representations by a ``contrastive'' loss which pushes apart dissimilar pairs while pulling together similar pairs. Majority of the CL related works focus on CV and models achieve significantly better performance than other approaches \cite{cl2015,cl2018,cl2019}. While some paid attention to learning robust sentence representations with CL by applying various augmentation strategies \cite{clnlp2020clear,clnlp2020declutr}. Few works applied CL to exercise representation learning.

\section{METHOD}

\subsection{Overall Framework}

\begin{figure}[ht]
\centering
\includegraphics[width=1.0\linewidth]{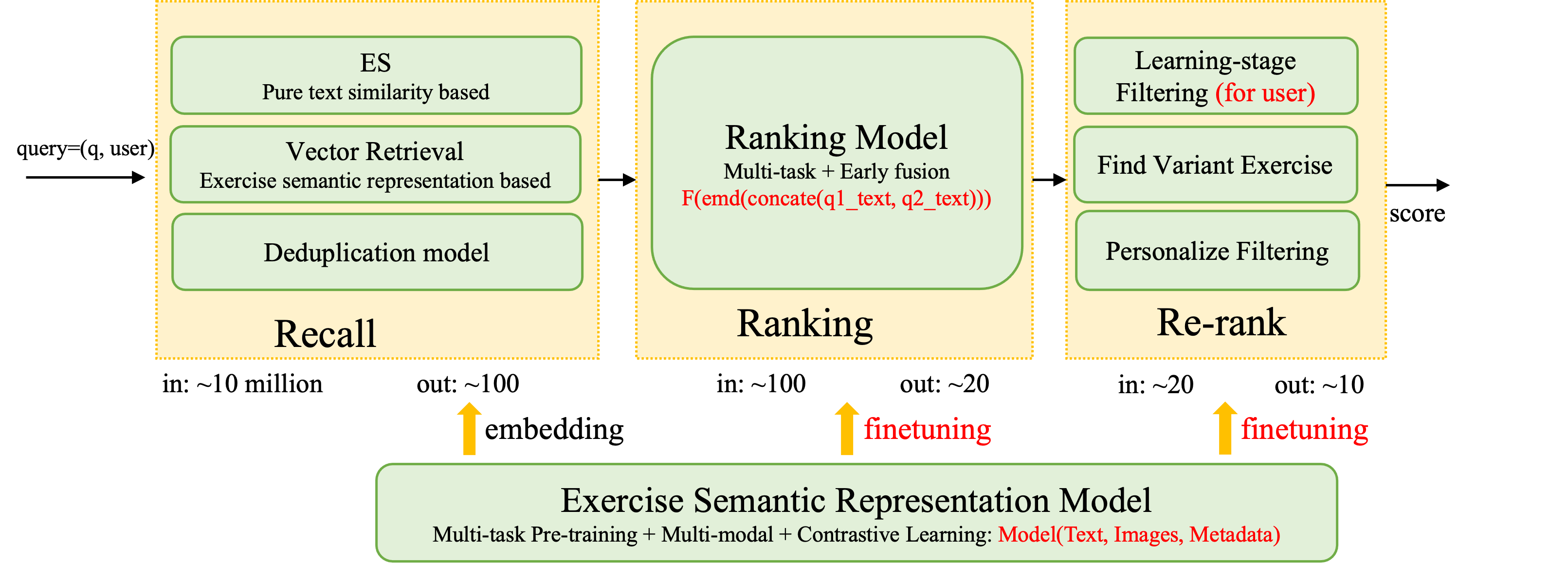}
\caption{Overall framework of our method.}
\label{fig:overall-framework}
\hfill 
\end{figure}

As described in the previous sections, the FSE task was modeled as a general retrieval process with modules of recall, ranking and re-rank. All modules are based on a unified \textbf{exercise semantic representation model (ESRM)}, which is used to learn the semantics of multi-modal exercise. The re-rank module is designed to satisfy practical needs specific to educational scenarios and we will introduce the logic behind it in depth. The overall framework of our FSE engine is in Figure \ref{fig:overall-framework} and details would be introduced next.

\subsection{Exercise Semantic Representation Module}
\label{repr-module}

In this section, we describe how to train an effective exercise representation model that is beneficial for all the recall, ranking, re-rank stages of our FSE engine. Although some works have been done to learn exercise representations, they are not targeted at FSE tasks or the effects are not ideal in FSE scenario. 

It is challenging to learn effective exercise representation for the FSE task. Firstly, the annotation of similar exercises is expensive and time-consuming. The definition of similar exercises is also subjective, which makes it difficult to obtain large amount of high-quality labeled data. With the limited noisy labeled data to supervise the training of exercise representation model, the risk of overfitting increases substantially. Secondly, as mentioned earlier, exercises usually comprise of multi-modal data including text, images and metadata which is noisy (eg. some labels in the metadata are incomplete and incorrect). To obtain effective representations, it is essential to take all the modalities into consideration while mitigating the noises. Thirdly, unlike general unstructured text, exercise is usually composed of stem, option, answer, and analysis. These structural information is also crucial for effective representation.

To overcome the above challenges, we propose the \textbf{general text pre-training (GTP)} $\rightarrow$ \textbf{multi-modal multi-task pre-training (MMP)} $\rightarrow$ \textbf{supervised fine-tuning (SFT)} learning approach. We adopt transformer as the backbone encoder which is shared through the three steps. In the general text pre-training step, encoder is trained on large amounts of unstructured text to obtain basic text semantic understanding capabilities. Subsequently, in the multi-modal multi-task pre-training step, the characteristics of the exercise text, the relationship among the exercise components, and the information contained in metadata and images are learned. Finally, we continue to fine-tune the model on the small amount of manually labeled similar exercises data. The first two steps can help to alleviate overfitting problem. The multi-modal multi-task pre-training is beneficial to cope with the latter two challenges.

\subsubsection{Data Pre-processing}
\label{data-process}
Before diving into the exercise representation learning method, we first introduce how to pre-process the multi-modal data.

\textbf{Exercise Text.}
\label{s400}The exercise text is cleaned by removing the stop words, HTML and CSS tags. Then we utilize the syntax parsing tool ANTLR\footnote{ANTLR (ANother Tool for Language Recognition): \\https://github.com/antlr/antlr4} to normalize the mathematical formula in the exercise text. The normalized exercise text is tokenized with tokenizer as in BERT-WWM\cite{chinese-bert-wwm} before feeding into the model. Each word in vocabulary is mapped into the vector representation with word embedding layer.

\textbf{Images.}
The chart, formula, illustrations, etc. may be in the form of images. We process the images as in QuesNet \cite{yin2019quesnet}. The images are converted to vector representations using convolutional network, which is then pretrained with auto-encoder loss.

\textbf{Metadata.} We use the exercise type, difficulty level and knowledge concepts in the metadata to pre-train the representation model. The exercise type and difficulty level are one-hot encoded. However, an exercise may have multiple different knowledge concepts. We encode knowledge concepts as normalized multi-hot vector $(\cdots, \frac{1}{n}, \cdots, \frac{1}{n}, \cdots)$ for exercise containing $n$ knowledge concepts.

\subsubsection{General Text Pre-training (GTP)}
Inspired by the strong semantic understanding ability of pre-training models which trained on large scale corpora, we train the transformer encoder on large amounts of unstructured text as a starting point for the subsequent multi-modal pre-training step.

\subsubsection{Multi-modal Multi-task Pre-training (MMP)}
\begin{figure}[ht]
\centering
\includegraphics[height=0.5\linewidth]{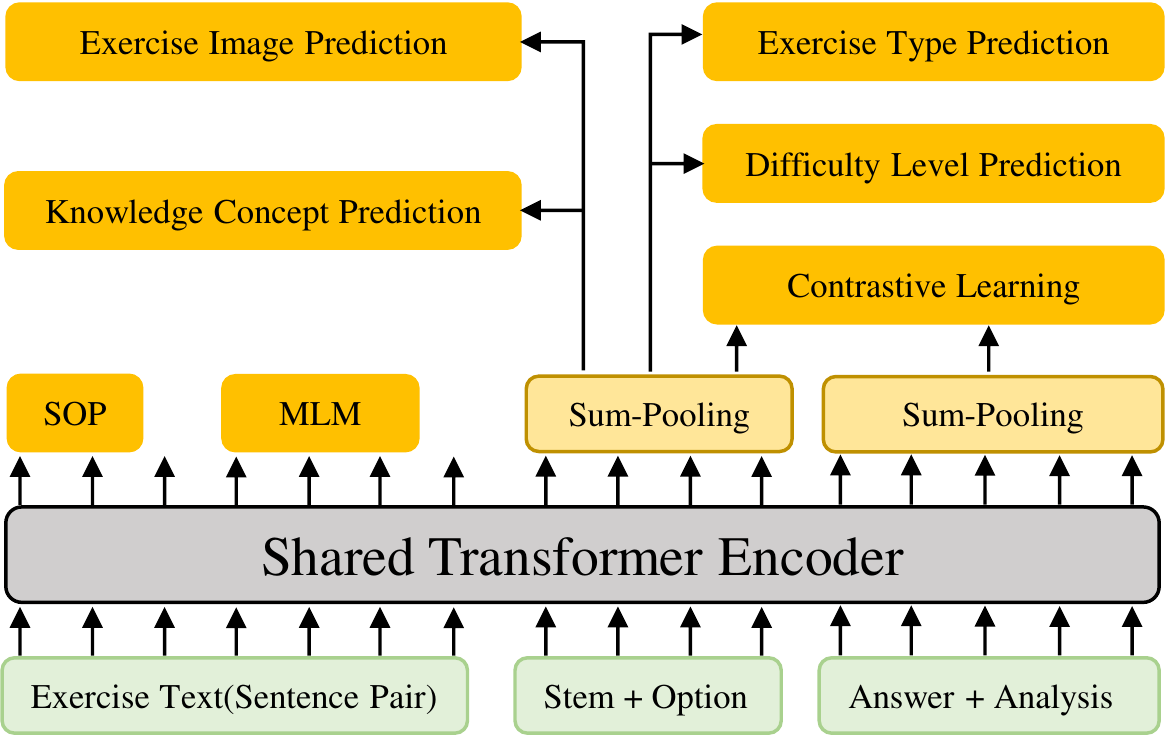}
\caption{Framework of multi-modal multi-task pre-training.}
\label{fig:multi-modal-pretrain}
\hfill 
\end{figure}

The framework of the multi-modal multi-task pre-training is shown in Figure \ref{fig:multi-modal-pretrain}. We will elaborate the pre-training tasks.

\textbf{SOP \& MLM.} Sentence order prediction (SOP) \cite{lan2019albert} and masked language model (MLM) are widely adopted in the BERT \cite{devlin2018bert} series models. Intuitively, SOP can be used to learn causal relationships in exercise text while MLM is beneficial for enhancing the representation of subject-specific terminology.

\textbf{Contrastive Learning.}
This task is to learn the relations between \emph{stem + option} and  \emph{answer + analysis} with the contrastive learning approach. The stem and option text, answer and analysis text of exercise $E_i$ are concatenated respectively and then fed into the encoder separately. We denote the top layer hidden states as $(h_{i_1}^s, h_{i_2}^s, ... , h_{i_m}^s)$ and $(h_{i_1}^a, h_{i_2}^a, ... , h_{i_n}^a)$ respectively, where $m$ is the length of \emph{stem + option} while $n$ is the length of \emph{answer + analysis}. The vector representations of \emph{stem + option} and  \emph{answer + analysis} of exercise $E_i$ are denoted as $e_i^{s}$ and $e_i^{a}$. They are obtained by the following formula:
\begin{equation}
    e_i^s = Norm(\sum_{k=1}^mh_{i_k}^s), \quad e_i^a = Norm(\sum_{k=1}^nh_{i_k}^a)
\end{equation}
where $Norm$ represents L2 normalization.

The objective of the contrastive learning is to make $e_i^{s}$ and $e_i^{a}$ more similar than $e_i^{s}$ and $e_j^{a}$, where $e_j^{a}$ is vector representation of \emph{answer + analysis} of exercise $E_j$ st. $i\not=j$. We adopt cosine similarity function to measure the similarity between the vectors. The loss for the task is expressed as:
\begin{equation}
    \mathcal{L}_{contrastive} = -\sum{\log_{}{\frac{e_i^{s} \cdot e_i^{a}}{e_i^{s} \cdot e_i^{a} + \sum_{j\not=i}{e_i^{s} \cdot e_j^{a}}}}}
\end{equation}
where exercise $E_j$ is a negative example that randomly sampled from the exercise bank.

\textbf{Exercise Type \& Difficulty Level \& Knowledge Concept Prediction.} The objective of these tasks is to classify the labels in metadata correctly. We stack a classification head upon the vector representation of \emph{stem + option} for each of the tasks. The training labels are in the form mentioned in Section \ref{data-process}. Cross entropy loss is adopted for the classifications. Predicting labels in metadata with the exercise text rather than feeding them into the encoder directly as feature is advantageous. On one hand, as mentioned earlier some labels in metadata are incorrect and incomplete. For example, completely labeling all knowledge concepts for some exercises is difficult. Adopting this approach not only allows the model to learn the information in metadata efficiently, but also mitigates the negative effects introduced by the imperfect labels. On the other hand, new exercise usually has no manually annotated metadata. The generalization ability of the model can be guaranteed in this way.

\textbf{Exercise Image Prediction.} We denote the image vector representations of exercise $E_i$ containing $p$ images as $(e_{i_1}^I, e_{i_2}^I, ... , e_{i_k}^I , ... , e_{i_p}^I)$, where $e_{i_k}^I$ is obtained with the method described in Section \ref{data-process}. We convert $e_{i_k}^I$ to $h_{i_k}^I$ with the following formula:
\begin{equation}
    h_{i_k}^I = Norm(Linear(e_{i_k}^I))
\end{equation}
where $Linear$ represents linear transform and $Norm$ means L2 normalization. The dimension of $h_{i_k}^I$ is the same as $e_i^{s}$.

The loss is same as the contrastive learning task, except we replace $e_i^{a}$ with $h_{i_k}^I$ and $e_j^{a}$ with $h_{j_k}^I$ where $h_{j_k}^I$ is from randomly sampled exercise $E_j$ st. $i \not= j$. The objective is to pull together $e_i^s$ and $h_{i_k}^I$ while push away $e_i^s$ and $h_{j_k}^I$.

Finally, the overall loss is the weighted sum of the losses of the several pre-training tasks.

\subsubsection{Supervised Fine-tuning (SFT)}
\label{fine-tune-representation}
We continue to fine-tune the pre-trained encoder on manually annotated pairs of similar exercises and adopt the contrastive learning approach again. Positive pairs are the labeled  similar exercises while negative pairs are randomly sampled from the exercise bank. Pilot experiments show that more complex negative sampling strategies (eg. hard negative sampling) is no better than the random sampling strategy in this scenario. The loss is denoted as:
\begin{equation}
    \mathcal{L}_{supervise} = -\sum{\log_{}{\frac{e_i^s \cdot e_{i+}^s}{e_i^s \cdot e_{i+}^s + \sum_{ \{ i- \}}{e_i^s \cdot e_{i-}^s}}}  }
\end{equation}
where exercise $E_{i+}$ is similar with exercise $E_i$ and exercise $E_{i-}$ is randomly sampled from the large exercise bank and dissimilar with exercise $E_i$.

Compared to directly applying multi-task training method, where the supervised fine-tuning task (FSE task) should be trained jointly with the auxiliary pre-training tasks (the tasks in the previous multi-task pre-training step), our \textbf{\emph{the pre-training then fine-tuning}} approach has advantages. Firstly, jointly training FSE task and the auxiliary tasks makes the whole process overly complex and inflexible. Secondly, we can obtain exercise embedding representation in an unsupervised manner with the current approach. That is, exercise embedding representation computed by the multi-task pre-trained encoder, which also can be used in other scenarios. 

\subsection{Recall Module}
\label{recall-module}
A few hundreds similar exercises are retrieved from millions of exercises in the recall module. We adopt two different methods to recall similar exercises separately. Then the similar exercises are merged with rules and detected duplicate exercises are removed. The first method is based on exact text matching to retrieve exercises with the highest degree of textual overlap. The second method is based on the exercise vector representation, which is used to retrieve exercises that are semantically related to the given exercise. The two methods are complementary. The exact text matching lacks the understanding of the semantics of the exercise while the vector representation based method is weak in exact text matching.

\subsubsection{Merging the Exercises}
We aim to select the most similar $n$ exercises from the exact text matching denoted as $E_{exact}$ and embedding matching denoted as $E_{embed}$ in the merging step. We adopt $E_{exact} \bigcap E_{embed}$ with its cardinality denoted as $n_u$, the most similar $\dfrac{n-n_u}{2}$ exercises from $E_{exact}\setminus E_{embed}$ and the most similar $\dfrac{n-n_u}{2}$ exercises from $E_{embed}\setminus E_{exact}$. 

\subsubsection{Duplicate Exercise Detection}

The duplicate exercise detection in the FSE engine is to improve the user experience in the consolidation exercise and avoid the repetitive homework. 
We adopt the model \textbf{ESRM} mentioned in Section \ref{repr-module} stacked with a binary classifier head to distinguish the duplicate pairs.

\subsection{Ranking Module}
\label{rank_module}

Dozens of similar exercises are ranked in the ranking module. There are still many challenges inherent in designing an effective ranking model such as the scarcity and in-sufficient semantically understanding of labeled exercises, and even worse the labeled pairs contains a lot of noise. 

\subsubsection{Multi-Task Learning for Insufficient Semantically Understanding}
\label{s42}
\begin{figure}[ht]
\centering
\includegraphics[height=0.6\linewidth]{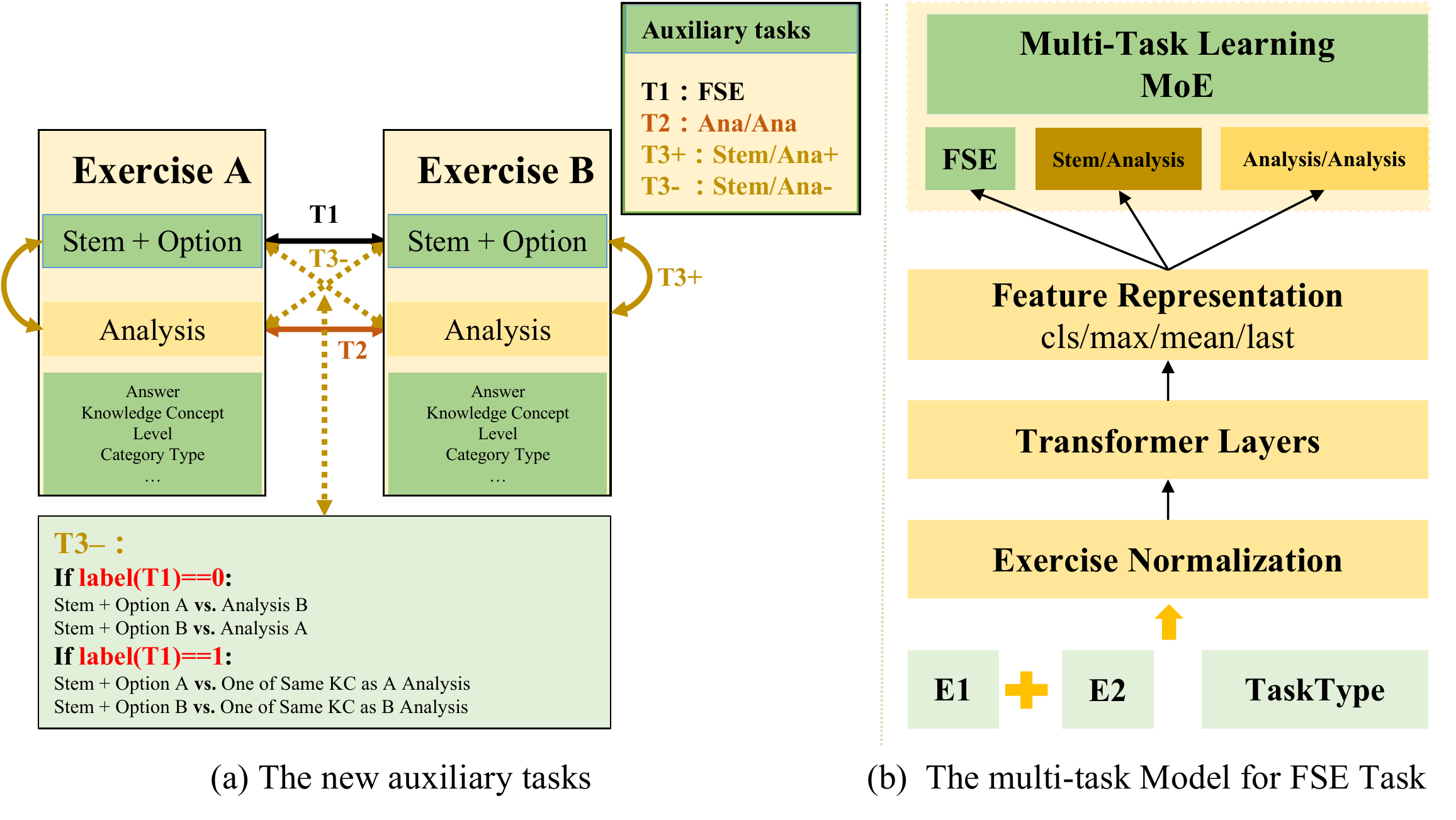}
\caption{(a) The way to find the new auxiliary tasks for FSE Task. (b) The framework of the ranking module.}
\label{fig:f4}
\hfill 
\end{figure}

Normal semantic understanding of similar exercises, which only depends on stems and options is not enough. In fact, the assessment of similar exercises not only needs to understand the exercises, but also needs to know how to solve the exercises. Therefore, how to assist the assessment of similar exercises with the help of problem-solving ideas? We find the exercise analysis which reflects the problem-solving ideas of exercise. Here, we discover new auxiliary tasks T2 and T3 in an innovative way by using the structural attribute of the exercise in Figure \ref{fig:f4}(a).  

\begin{table*}[ht!]
  \caption{The performance of the exercise representation learning methods. ESRM=GTP+MMP+SFT.}
\centering
    \begin{tabular}{l|r|lrc|ccccc}
    \toprule
        Model & ETM & MANN & SBERT & QuesNet & GTP+MMP & GTP+SFT & MMP+SFT & ESRM & ESRM+ETM \\ \hline
        R@100 (recall) & 72.49\% & 63.64\% & \textbf{73.51}\% & 71.32\% & 60.84\% & 79.67\% & 82.81\% & \textbf{86.48}\% & \textbf{88.94}\% \\ \hline
        P@5 (ranking)   & 71.45\% & 68.57\% & \textbf{75.91}\% & 73.75\% & 82.34\% & 76.92\% & 80.59\% & \textbf{83.83}\% & -  \\ \hline
    \end{tabular}
    \label{table:representation-learning}
\end{table*}

\subsubsection{Discovering New Tasks}
\label{s43}

Taking a labeled tuple (Exercise A, Exercise B, label) as an example, the stem and options of exercise A and exercise B constitute the FSE task T1. Exercise analysis A and exercise analysis B constitute a new task T2 while the stem and analysis can be constructed a new task T3. Specifically, the positive pair can be directly constructed with the original stem and analysis(T3+), while the negative pair is slightly different that depends on the label of exercise pair. If the label is dis-similar, we can directly construct the stem A and analysis B. If the label is similar, directly construct may bring in noise. Here, we regard the current exercise stem and the same concepts of knowledge with randomly select an exercise analysis as a negative example (T3-).

\subsubsection{MoE enhanced Multi-Task Learning}
\label{ranking:moe}

Following the work \cite{Liu2019multi}, we adopt the hard-sharing multitask learning in Figure \ref{fig:f4}(b) which the bottom layer shares the same encoder(\textbf{ESRM} model) and the upper task layer learns different task representations. The multi-task model can learn more logical and semantic representation from exercise analysis and decrease the occurrence of over-fitting. The query exercise and candidate exercise are processed mentioned in Section \ref{data-process} and concatenated as the input of \textbf{ESRM} model. Then the exercise feature representation is extracted by some pooling methods such as sum or mean. Finally, for each task, additional task-specific layers generate task-specific representations, followed by operations necessary for specific task. The overall loss of multi-task is expressed as follows:
\begin{equation}
 \mathcal{L} = \alpha_{1}\cdot \mathcal{L}_{T1} + \alpha_{2} \cdot \mathcal{L}_{T2}+ \alpha_{3} \cdot \mathcal{L}_{T3}
\label{e1}
\end{equation}
where $\mathcal{L}_{T1}$ represents the loss of FSE task, $ \mathcal{L}_{T2}$ is the match loss of analysis, $ \mathcal{L}_{T3}$ is the match loss of stem and analysis, $\alpha_i$ is the task coefficient which st. $\sum_{i=1}^{n=3}(\alpha_i)=1$.Since adjusting the task coefficient depends on manual attempts and expert experience, we adopt a 3-layer neural network to \textit{dynamically learn} the coefficients of tasks which is similar to MoE Layer \cite{jacobs1991adaptive,ma2018modeling} in information recommendation field.

\subsubsection{Confidence Learning}
\label{s41}

Due to the human labeled education data requires strong expertise, the differences exist in teachers' teaching experience and understanding of FSE problem. The statistical results show that the consistency rate of labeling samples is usually between 80-85\%, so there are some noises in the labeled samples. As we all know, if there is too much noise, the model will be easier to fit to the noise samples. In order to prune and denoise the labeled samples, we introduce the confidence learning \cite{northcutt2021confident} to resolve this problem. The application of confidence learning can remove some noisy samples in theory and obtain a better performance.

\subsection{Re-rank Module}
\label{rerank_module}

The re-rank technique is generally used as a post-processing step to boost the user experience in various retrieval or recommendation problems \cite{xu2018deep}. After we obtain the the ranking exercise list, we need refine the ranking list to meet the user expected exercises in the FSE engine. The re-rank module includes the following parts: variant exercise re-rank, personalized exercise filtering and learning stage filtering.

\section{EXPERIMENTS}

\begin{table*}[ht!]
  \caption{Overall performance of development dataset results.}
\centering
    \begin{tabular}{l|rrr|lrrc|c}
    \toprule
        Model & OVSM & BERT & SBERT  & ESRM & +MTL & +MoE & +ConL & $\Delta$(vs. max(OVSM, BERT, SBERT)) \\ \hline
        P@1 & 80.68\% & 75.81\% & \textbf{82.80\%} & 88.62\% & 90.02\% & 91.22\% & \textbf{92.68\%} & \textbf{+9.88\%$\uparrow$} \\ \hline
        P@3 & \textbf{79.64\%} & 74.23\% & 78.72\%  & 85.01\% & 88.23\% & 89.62\% & \textbf{90.59\%} & \textbf{+10.95\%$\uparrow$} \\ \hline
        P@5 & \textbf{78.99\%} & 70.22\% & 75.91\% & 83.83\% & 85.57\% & 86.77\% & \textbf{89.14\%} & \textbf{+10.15\%$\uparrow$} \\\hline
    \end{tabular}
    \label{table:t1}

\end{table*}

\subsection{Offline Experiments}
We conduct experiments based on the offline dataset (detailed in section \ref{sb1}) to evaluate the recall module with R@100 and the ranking module with P@K(1/3/5). 

\subsubsection{Exercise Representation Module}
With same settings as our implementation, we choose the mainstream exercise representation learning methods(MANN \cite{liu2018finding}, SBERT and QuesNet) to compare. 

\textbf{Results.}
We can observe that our exercise semantic representation model (\textbf{ESRM}) performs the best among all the compared methods when applied to both recall and ranking stages in the Table \ref{table:representation-learning}. Combining with the extract text matching method and our vector representation based method can further improve the R@100 metric significantly, which indicates that the two methods are complementary, as mentioned earlier. By removing either of the training phases in our exercise representation learning model, the performance drops evidently which shows that each of the training phases plays an important role for learning effective exercise representation.

\begin{figure}[ht]
\centering
\includegraphics[height=0.8\linewidth]{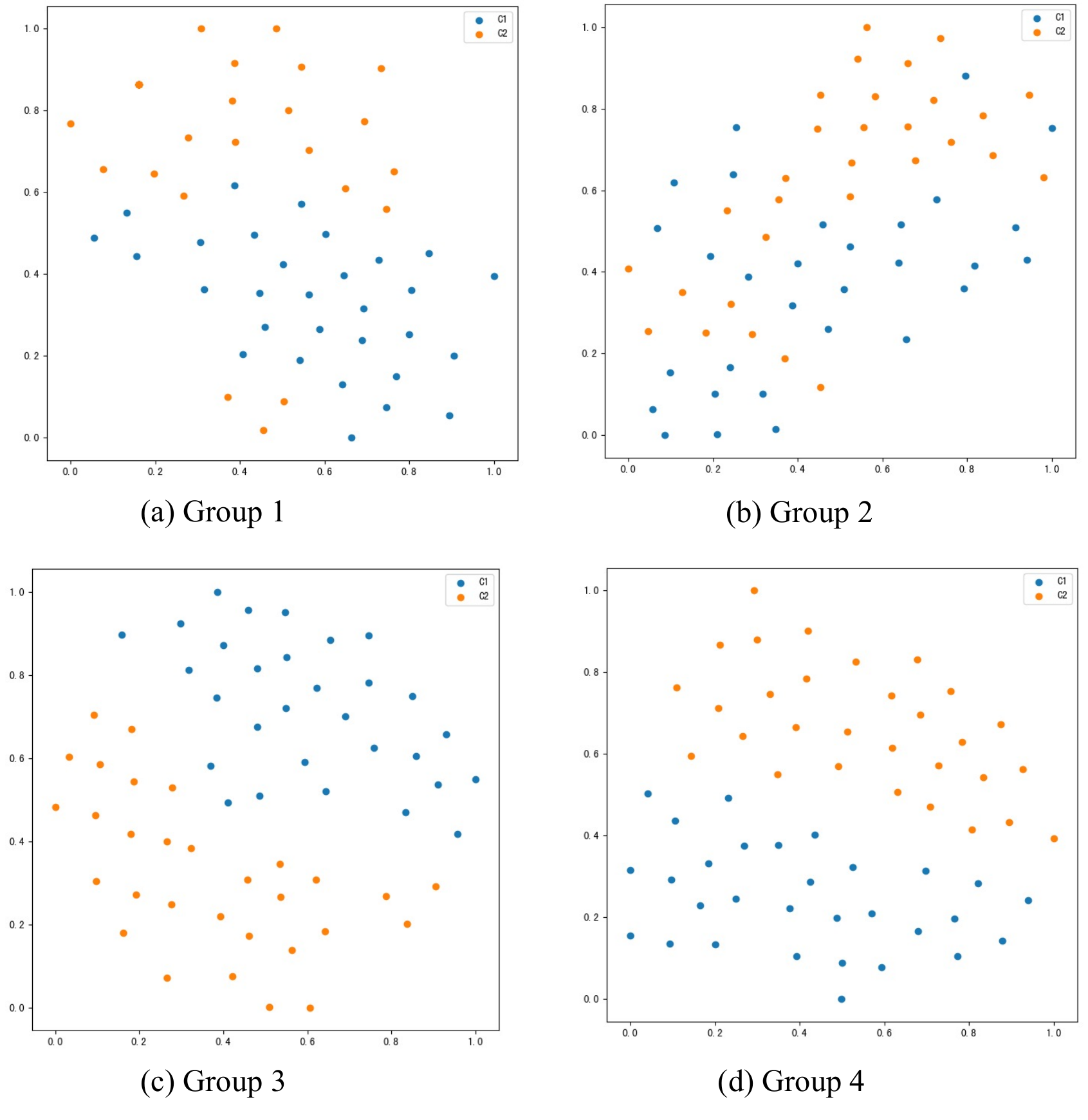}
\caption{Visualization of exercise representation. The knowledge concepts of C1 and C2 in Figure(a) and Figure(b) are relatively similar, while C1 and C2 in Figure(c) and Figure(d) are quite different.}
\label{fig:vis}
\hfill 
\end{figure}

\subsubsection{Ranking Module}
We compared with some benchmark methods (OVSM \cite{yu2014similarity}, BERT-Base \cite{devlin2018bert} and SBERT \cite{feng2019learning}) on junior mathematics data set (detailed in section \ref{sb1}). 

\textbf{Results.}
We have the following observations in Table \ref{table:t1}:
\textbf{\\1.} The (ESRM) is able to boost the performance and consistently outperforms other baseline models. This proves that our model gains a better understanding of exercisers and is more efficiently to transfer from large unlabeled corpus to the label-scarce dataset.
\textbf{\\2.} The experimental results also confirm the effectiveness of multi-task learning (+MTL) and provide more high level logic information for FSE task. In addition, the experimental results show that MoE mechanism (+MoE) is very effective.
\textbf{\\3.} Confidence Learning (+ConL) aims to prune train-set and gains further improvement.

\subsection{Visualization of the Exercise Representation}

We conduct visualization analysis of the exercise representation module. It is important to learn exercise representations in which similar exercises are closer while dissimilar exercises are farther. To show the results intuitively, we first select four groups of exercises under two concepts that are randomly selected (the first two groups have the close knowledge concepts while the latter two are different), and then reduce the dimension of obtained representations by t-SNE \cite{van2008visualizing}. The visualization results are shown in Figure \ref{fig:vis} and we can get two interesting phenomena. On one hand, if the knowledge concepts of C1 and C2 are relatively similar (Fig \ref{fig:vis}a and Fig \ref{fig:vis}b), their exercises representation are also relatively close. There may be some overlapping parts because similar exercises are those having a similar logic and relationships including not only knowledge concepts but also some other information such as problem-solving ideas. On the other hand, if the knowledge concepts of C1 and C2 are quite different (Fig \ref{fig:vis}c and Fig \ref{fig:vis}d), they are unlikely to become similar exercises and their exercises representation have a big difference. Thus, the results show that our exercise representation seems good intuitively.

\subsection{Online Serving and A/B Test}
\label{online_exp}

Motivated by the encouraging offline evaluation results, we deploy the FSE engine in Tencent education platform and test its performance online. From the results of A/B test, the ranking model based on MTL+MoE+ConL gains \textbf{15.98\%} improvement for user acceptance rate and precision, while the retrieval method based on ESRM in recall module gains \textbf{3.9\%} improvement further, compared with an OVSM \cite{yu2014similarity} based early version. The FSE engine brings the improvement of user satisfaction and teaching efficiency.

It is worth noticing that online FSE system is of great challenge for serving hundreds of schools. To keep low latency and high throughput, we deploy several important strategies to improve performance: 1) combining the offline cache with online system to avoid repeated calculation, 2) Batching: adjacent user requests are merged into one batch to take advantage of GPUs, 3) The request from the same user is sent in parallel to each module.

\section{CONCLUSION AND FUTURE WORK}

This paper proposes a personalized similar exercise retrieval system which is effective in production environment, rather than a specific algorithm. In terms of module design, this system is roughly the same as the existing retrieval or recommendation systems. The differences lie in that the characteristics of the FSE task and specific business needs are taken into consideration within each module. All the models such as the multi-modal exercise representation model, the multi-task based ranking model and the confidence learning for handling noise data are optimized for the FSE scenario. The paper also elaborates the rationality of the model architecture design. In the future work, we would try to introduce graph neural network to represent the multi-modal exercise data and design personalized recall and ranking models to make similar exercises more personalized and satisfactorily.

\bibliographystyle{aaai23}
\bibliography{aaai23} 

\begin{thebibliography}{22}
\providecommand{\natexlab}[1]{#1}

\bibitem[{Cui et~al.(2019)Cui, Che, Liu, Qin, Yang, Wang, and
  Hu}]{chinese-bert-wwm}
Cui, Y.; Che, W.; Liu, T.; Qin, B.; Yang, Z.; Wang, S.; and Hu, G. 2019.
\newblock Pre-Training with Whole Word Masking for Chinese BERT.
\newblock \emph{arXiv preprint arXiv:1906.08101}.

\bibitem[{Devlin et~al.(2018)Devlin, Chang, Lee, and
  Toutanova}]{devlin2018bert}
Devlin, J.; Chang, M.-W.; Lee, K.; and Toutanova, K. 2018.
\newblock Bert: Pre-training of deep bidirectional transformers for language
  understanding.
\newblock \emph{arXiv preprint arXiv:1810.04805}.

\bibitem[{Doersch, Gupta, and Efros(2015)}]{cl2015}
Doersch, C.; Gupta, A.; and Efros, A.~A. 2015.
\newblock Unsupervised visual representation learning by context prediction.
\newblock \emph{ICCV}.

\bibitem[{Feng et~al.(2019)Feng, Chen, Guo, Zhao, and Fu}]{feng2019learning}
Feng, M.; Chen, Y.; Guo, Y.; Zhao, Y.; and Fu, G. 2019.
\newblock Learning Text Representations for Finding Similar Exercises.
\newblock In \emph{2019 IEEE International Conference on Consumer
  Electronics-Taiwan (ICCE-TW)}, 1--2. IEEE.

\bibitem[{Giorgi et~al.(2020)Giorgi, Nitski, Wang, and
  Bader}]{clnlp2020declutr}
Giorgi, J.; Nitski, O.; Wang, B.; and Bader, G. 2020.
\newblock Declutr: Deep contrastive learning for unsupervised textual
  representations.
\newblock \emph{arXiv preprint arXiv:2006.03659}.

\bibitem[{Jacobs et~al.(1991)Jacobs, Jordan, Nowlan, and
  Hinton}]{jacobs1991adaptive}
Jacobs, R.~A.; Jordan, M.~I.; Nowlan, S.~J.; and Hinton, G.~E. 1991.
\newblock Adaptive mixtures of local experts.
\newblock \emph{Neural computation}, 3(1): 79--87.

\bibitem[{Komodakis and Gidaris(2018)}]{cl2018}
Komodakis, N.; and Gidaris, S. 2018.
\newblock Unsupervised representation learning by predicting image rotations.

\bibitem[{Lan et~al.(2019)Lan, Chen, Goodman, Gimpel, Sharma, and
  Soricut}]{lan2019albert}
Lan, Z.; Chen, M.; Goodman, S.; Gimpel, K.; Sharma, P.; and Soricut, R. 2019.
\newblock Albert: A lite bert for self-supervised learning of language
  representations.
\newblock \emph{arXiv preprint arXiv:1909.11942}.

\bibitem[{Lee, Na, and Kim(2019)}]{lee2019multi}
Lee, W.; Na, J.; and Kim, G. 2019.
\newblock Multi-task self-supervised object detection via recycling of bounding
  box annotations.
\newblock In \emph{Proceedings of the IEEE/CVF Conference on Computer Vision
  and Pattern Recognition}, 4984--4993.

\bibitem[{Liu et~al.(2018)Liu, Huang, Huang, Liu, Chen, Su, and
  Hu}]{liu2018finding}
Liu, Q.; Huang, Z.; Huang, Z.; Liu, C.; Chen, E.; Su, Y.; and Hu, G. 2018.
\newblock Finding similar exercises in online education systems.
\newblock In \emph{Proceedings of the 24th ACM SIGKDD International Conference
  on Knowledge Discovery \& Data Mining}, 1821--1830.

\bibitem[{Liu et~al.(2019)Liu, He, Chen, and Gao}]{Liu2019multi}
Liu, X.; He, P.; Chen, W.; and Gao, J. 2019.
\newblock Multi-task deep neural networks for natural language understanding.
\newblock \emph{arXiv preprint arXiv:1901.11504}.

\bibitem[{Ma et~al.(2018)Ma, Zhao, Yi, Chen, Hong, and Chi}]{ma2018modeling}
Ma, J.; Zhao, Z.; Yi, X.; Chen, J.; Hong, L.; and Chi, E.~H. 2018.
\newblock Modeling task relationships in multi-task learning with multi-gate
  mixture-of-experts.
\newblock In \emph{Proceedings of the 24th ACM SIGKDD International Conference
  on Knowledge Discovery \& Data Mining}, 1930--1939.

\bibitem[{Northcutt, Jiang, and Chuang(2021)}]{northcutt2021confident}
Northcutt, C.; Jiang, L.; and Chuang, I. 2021.
\newblock Confident learning: Estimating uncertainty in dataset labels.
\newblock \emph{Journal of Artificial Intelligence Research}, 70: 1373--1411.

\bibitem[{Pardos and Dadu(2017)}]{pardos2017imputing}
Pardos, Z.~A.; and Dadu, A. 2017.
\newblock Imputing kcs with representations of problem content and context.
\newblock 148–155.

\bibitem[{Peng and Dredze(2016)}]{peng2016improving}
Peng, N.; and Dredze, M. 2016.
\newblock Improving named entity recognition for chinese social media with word
  segmentation representation learning.
\newblock \emph{arXiv preprint arXiv:1603.00786}.

\bibitem[{Van~der Maaten and Hinton(2008)}]{van2008visualizing}
Van~der Maaten, L.; and Hinton, G. 2008.
\newblock Visualizing data using t-SNE.
\newblock \emph{Journal of machine learning research}, 9(11).

\bibitem[{Wang, Che, and Manning(2013)}]{wang2013joint}
Wang, M.; Che, W.; and Manning, C.~D. 2013.
\newblock Joint word alignment and bilingual named entity recognition using
  dual decomposition.
\newblock In \emph{Proceedings of the 51st Annual Meeting of the Association
  for Computational Linguistics (Volume 1: Long Papers)}, 1073--1082.

\bibitem[{Wu et~al.(2020)Wu, Wang, Gu, Khabsa, Sun, and Ma}]{clnlp2020clear}
Wu, Z.; Wang, S.; Gu, J.; Khabsa, M.; Sun, F.; and Ma, H. 2020.
\newblock Clear: Contrastive learning for sentence representation.
\newblock \emph{arXiv preprint arXiv:2012.15466}.

\bibitem[{Xu, He, and Li(2018)}]{xu2018deep}
Xu, J.; He, X.; and Li, H. 2018.
\newblock Deep learning for matching in search and recommendation.
\newblock In \emph{The 41st International ACM SIGIR Conference on Research \&
  Development in Information Retrieval}, 1365--1368.

\bibitem[{Yin et~al.(2019)Yin, Liu, Huang, Chen, Tong, Wang, and
  Su}]{yin2019quesnet}
Yin, Y.; Liu, Q.; Huang, Z.; Chen, E.; Tong, W.; Wang, S.; and Su, Y. 2019.
\newblock QuesNet: A Unified Representation for Heterogeneous Test Questions.
\newblock In \emph{Proceedings of the 25th ACM SIGKDD International Conference
  on Knowledge Discovery \& Data Mining}, 1328--1336.

\bibitem[{Yu et~al.(2014)Yu, Li, Hou, Liu, and Yang}]{yu2014similarity}
Yu, J.; Li, D.; Hou, J.; Liu, Y.; and Yang, Z. 2014.
\newblock Similarity Measure of Test Questions Based on Ontology and VSM.
\newblock \emph{The Open Automation and Control Systems Journal}, 6(1).

\bibitem[{Zhang et~al.(2019)Zhang, Qi, Wang, and Luo}]{cl2019}
Zhang, L.; Qi, G.-J.; Wang, L.; and Luo, J. 2019.
\newblock Aet vs. aed: Unsupervised representation learning by auto-encoding
  transformations rather than data.
\newblock 2547--2555.

\end{thebibliography}

\appendix

\clearpage

\section{APPENDIX}

\subsection{Problem Formulation}
\label{s3}
As mentioned earlier, similar exercises are those having the same purpose which related to the semantics of exercises.

\begin{definition}
Given a set of exercises including stem, option, answer and analysis, our target is to learn a model $\mathcal{F}$ which can be used to measure the similarity scores of exercise pairs and find similar exercises for any exercise E by ranking the candidate ones $\mathcal{D}$ with similarity scores:
\begin{equation}
    \mathcal{F}(E, \mathcal{D}, \Theta) \to \mathcal{R}^{s}
\end{equation}
where $\Theta$ is the parameters of $\mathcal{F}$, $\mathcal{D}=(E_1, E_2, E_3, \cdots)$ are the candidate exercises for $E$ and $\mathcal{R}^{s} = (E_1^s, E_2^s, E_3^s, \cdots)$ are the candidates ranked in descending order with their similarity scores $(S(E,E_1^s),S(E,E_2^s),S(E,E_3^s),\cdots)$. The similar exercises for E are those candidates having the largest similarity scores.
\end{definition}

\section{Experimental Setup}

\subsection{Dataset}
\label{sb1}
The Tencent education platform\footnote{Tencent education platform: https://edu.tencent.com} contains millions of exercises and we only choose about 350K junior math exercises for our experiments. We sample 1.5K seed exercises and construct 23K exercise pairs through the BM25 match and some strategies such as random choose and random with knowledge concepts. Then theses exercises are labeled with several similar exercises and each given exercise is labeled by three teachers. We choose the majority numbers of votes as the label for the similar exercise. We split our data set randomly via the seed exercises into three parts: 80\% is for training set, 10\% is for validation set and 10\% is for test set. Finally, we only report the performances on the test set.

\subsection{Metrics}

\label{sbm}

We describe the metrics we use to evaluate the recall module and the ranking module here.

\textbf{Recall Metrics.}
We focus on the metric Recall@K here. Recall@K measures the ratio of the annotated similar exercises of the seed exercise that are ranked at top K by the recall method. The metric is defined as:
\begin{equation}
\small
    Recall@K = \frac{\sum_{i=1}^{N}(TP_i@K \ \slash \ T_i)}{N}
\end{equation}
where $T_i$ is the number of annotated similar exercises of seed exercise $i$, $TP_i@K$ measures how many of the $T_i$ similar exercises are ranked at top K by the recall method and N is the number of annotated seed exercises. We take Recall@100 (abbr. \textbf{R@100}) in this paper.

\textbf{Ranking Metrics.}
For any \textit{exercise E}, we could find its similar exercises by ranking the candidate ones according to their similarity scores, and finally return the accurate top-K similar ones. We use the Precision@K to measure how many relevant exercises are present in the top-K with FSE engine. We take Precision@1/3/5 (abbr. \textbf{P@1/3/5}) in this paper.

\subsection{Exercise Semantic Representation Module}
\label{sb2}
\textbf{Experimental Setup.}
For pre-training our representation learning model, we follow BERT-WWM\cite{chinese-bert-wwm}. We set the weight for \textbf{SOP}, \textbf{MLM} and \textbf{Contrastive Learning} as 1.0 and that for the other pre-training tasks as 0.5. For the fune-tuning, we set learning rate as 1e-5 and train on data for 4 epoches with batch size of 128.

We implement exact text matching method with ElasticSearch\footnote{ElasticSearch: https://www.elastic.co/elasticsearch}. The exercise text and knowledge concepts are stored in separate fields of the engine and used for scoring. For the compared methods, when used for ranking, we replace our transformer encoder layers with their exercise representation encoder for a fair comparison. Annoy\footnote{Annoy: https://github.com/spotify/annoy} is used for searching the most similar exercises in the vector representation based method of the recall module.

\subsection{Duplicate Exercise}
\begin{table}[!hbt]
    \centering
    \small
    \caption{The pair examples of duplicate exercises.}
    \label{table:trerank}
    \begin{tabular}{l|l}
    \hline
         & Example \\ \hline
        E1.a & After the price of a commodity is increased by 25\%,\\ & if you want to restore the original price, you\\ & should reduce the price\_\_\_ \\ \hline
        E1.b & After the price of a commodity is increased by 25\%\\& in  May 2020, if the original price is to be restored\\ & in May 2021, the price shall be reduced\_\_\_ \\ \hline
        E2.a & Given that x = - 1 is a root of equation $x-2m=0$,\\ & the value of M is \_\_ \\ \hline
        E2.b & Given that x = - 1 is a root of equation $x^2-2m=0$,\\ & the value of M is \_\_ \\ \hline
    \end{tabular}
\end{table}

Taking the examples in Table \ref{table:trerank}, the original exercise E1.a is how much the price of goods is increased, in order to restore the original price and how much the price should reduced, while the duplicate exercise E1.b is to add some disturbing year digital information. Although there is only a numerical difference for E2.a and E2.b in example, it is a difference between uni-variate quadratic equation and uni-variate primary variance, which does not constitute a duplicate pair. 

\subsection{Ranking Module}
\label{sb3}
\subsubsection{Experimental Setup.}
We implement all the models with Tensorflow in our experiments. In the pre-training stage, we pre-train our models with MLM objective, continuing from the published checkpoint, BERT-base-chinese. We pre-train our model 
for 200K steps, and the first 3000 steps are for warm-up. The rest of the hyper-parameters are the same as BERT-base. In the fine-tuning stage, we train our model in the multi-task paradigm. The multi task module adopts three layers of neural network, and the output sizes of hidden layer of each layer are 768, 768 and 3. We apply the Adam method to optimize our model. The learning rate is $2e-5$, the number of training epoch is 3. We conduct our experiments with 2 Tesla T4 GPUs.

\subsubsection{The Detail Of MoE enhanced Multi-Task Learning}
The detail MoE operations are as follows:
First of all, we concat the feature representations of the different tasks:
\begin{equation}
\small
feature= concat(Fe_{T_{1}}, Fe_{T_{2}}, Fe_{T_{3}})
\label{e2}
\end{equation}
Secondly, for the feature representation, we learn the task of parameter coefficients through a three-layer neural network.
\begin{equation}
\alpha = (\alpha_{1},\alpha_{2},\alpha_{3}) = concat(g_1{(Fe_{T_{1}})},g_2{(Fe_{T_{2}})}, g_3{(Fe_{T_{3}})})
\label{e3}
\end{equation}
where $g_i(.)$ is the $i$-th expert network with a three-layer neural network. Finally, the total loss is the \textit{dynamically weighted} sum of the loss of the three tasks as the above Formula \ref{e1}.

\subsection{The Detail Of Confidence Learning}
The application of confidence learning to noise
data is mainly divided into three stages. The first stage is estimation
and estimates the joint probability distribution of noisy labels and
real labels, which relies on a model to predict the probability of
exercise pair. The second stage is pruning and uses the previously
calculated joint probability distribution for pruning according to
the level of confidence. The third stage is re-training based on the
cleaned data set.

\section{Re-rank Module}
\subsection{Variant Exercise re-rank}
\label{var_reranking}
The variant exercise refers to the exercise which is changed the condition, conclusion or the problem-solving method of the exercise. Compare with the regular consolidation exercises, it can improve students' different thinking and innovative consciousness more. Therefore, we need to divide the generalized similar exercise list into similar exercise list and variant exercise list. However, it is not easy to distinguish the variant exercise because of higher expertise knowledge, more expensive expenses, scarcity and lower consistency of labeling data. We use the exercise presentation model \textbf{ESRM} mentioned in Section \ref{repr-module} as a binary classifier to classify variant exercise and similar exercise. The final retrieval exercise list to users are those with high variant scores in the re-rank stage.

\subsection{Personalized Exercise Filtering}
The personalized exercise filtering is crucial to the final retrieval. The personalized exercise filtering depends on students' ability. If a student is an excellent student, then we recommend an exercise similar or hard in difficulty to this exercise for consolidation. Similarly, if a student is a weak student, we recommend an exercise similar or easy in difficulty to this exercise for consolidation.

\subsection{Learning Stage Filtering}
The students' learning stage filtering module will decide which exercises in the ranking list can be recommended or not according to the students' learning stage. If a student is in the synchronous practice stage, the exercises to be recommended should not exceed the current learning stage, and if a student is in the review stage, the exercises to be recommended should not exceed the current semester.

\end{document}